# An Empirical Comparison of Machine Learning Models for Student's Mental Health Illness Assessment

Prathamesh Muzumdar[1], Ganga Prasad Basyal[2], Piyush Vyas[3]

[1]College of Business, The University of Texas at Arlington, Texas, USA
*Email: prathameshmegh.muzumdar [AT] mavs.uta.edu*

[2]Business and Natural Sciences Department, College of Business and Natural Sciences
Black Hills State University, Spearfish, South Dakota, USA
*Email: gangaprasad.basyal [AT] bhsu.edu*

[3]College of Business and Information Systems, Dakota State University
Madison, South Dakota, USA
*Email: piyush.vyas [AT] trojans.dsu.edu*

_________________________________________________________________________________________

**ABSTRACT---** *Student's mental health problems have been explored previously in higher education literature in various contexts including empirical work involving quantitative and qualitative methods. Nevertheless, comparatively few research could be found, aiming for computational methods that learn information directly from data without relying on set parameters for a predetermined equation as an analytical method. This study aims to investigate the performance of Machine learning (ML) models used in higher education. ML models considered are Naïve Bayes, Support Vector Machine, K-Nearest Neighbor, Logistic Regression, Stochastic Gradient Descent, Decision Tree, Random Forest, XGBoost (Extreme Gradient Boosting Decision Tree), and NGBoost (Natural) algorithm. Considering the factors of mental health illness among students, we follow three phases of data processing: segmentation, feature extraction, and classification. We evaluate these ML models against classification performance metrics such as accuracy, precision, recall, F1 score, and predicted run time. The empirical analysis includes two contributions: 1. It examines the performance of various ML models on a survey-based educational dataset, inferring a significant classification performance by a tree-based XGBoost algorithm; 2. It explores the feature importance [variables] from the datasets to infer the significant importance of social support, learning environment, and childhood adversities on a student's mental health illness.*

## 1. INTRODUCTION

The World Health Organization [WHO] has indicated that mental illness affects nearly half of the population worldwide [3]. The ubiquity of mental illness is associated with considerable impairment in cognitive skills [13], with the combination of anxiety and affective disorders being the most common forms of disability among sufferers [20]. Mental illness is as prevalent among college students as others, and the illness appears to be increasing in number and severity [24]. When researchers study such illness, college students are the most neglected population, though students are not immune to the sufferings and disability associated with such illness. Mental health problems are very common among students and are highly prevalent in college students [14]. This is mainly because attending college requires a student to overcome challenging times, which is uncommon in settings like community college and high school [26]. College students start college after completing high school, typically depend upon parents for financial support for few years till they start working part-time [1]. These students not only face stress related to academic load but also have to face the task of taking on more adult-like responsibilities at a younger age. Mental health problems among college students represent a growing concern as a large number of students enter their adulthood and this stage is considered an important period of life [24]. The developmentally challenging transition to adulthood and untreated mental illness leads to crucial implications for academic success, productivity, social and personal relationships, and substance abuse [29].

College tenure represents the time in many people's lives when their main activities relating to career and social life are integrated with their surroundings. University campuses are responsible to develop, disseminate, and evaluate best practices to ensure a better experience for students [15]. Therefore, colleges offer the best opportunity to address student's mental health problems among late adolescents and young adults [13]. A robust base of research is necessary to deeply understand this phenomenon and investigate the antecedents of student's mental health problems. In that regard, machine learning models and data mining techniques are efficient methods in extracting features and classifying educational datasets [6]. Universities today are attempting to improve their performance and policy reforms by employing ML models and data mining techniques to educational datasets. The educational datasets containing information on student's mental health are often characterized by many factors (variables) that are responsible for developing mental illness among students. Such variables are considered as factors responsible for developing mental illness among students. Dataset with such a diverse characteristic makes it challenging to analyze and comprehend data. Prior research has focused on developing and analyzing conceptual models to contribute to the theoretical foundation of this research genre. Only a few past research have taken into consideration on developing and improving ML models for diverse dataset. Most research has limited their focus on either improving feature learning or finding significant techniques with the best speed and accuracy parameters. They have failed to account for a detailed approach that could have been implemented to understand the performance of the ML model and map them to the characteristic variables of the educational dataset.

This study aims to analyze the performance of different ML classification models using various datasets. As educational datasets vary with diverse characteristic variables, this diversity in variables has been evident to impact the ML model's hyperparameter tuning, classification performance, and run time [6]. This research further investigates the performance of ML classification models on educational data in the context of student's mental health illnesses. The significance of this research is both theoretical and practical. From a theoretical perspective, this research helps to understand the benefits and deficiencies of ML models that could help build a foundation for future research in improving the ML classification models for educational datasets. From a practical perspective, this research helps to understand the applied research application of ML classification models to educational datasets. Overall, this research tries to address the following questions in this study:

1. Which ML classification model best fits the student's mental health illness datasets?

2. Which features (variables) significantly influence student's mental health illness when considering all the datasets and best model fit?

The remainder of this paper is organized as follows: Section 2 discusses the theoretical background of the study, taking into consideration of the literature on student's mental health illness and comparative analysis of ML classification models. Section 3 discusses the methodology, which includes a brief discussion on the characteristics of datasets and details of the analysis process. Section 4 illustrates the results from the ML model analysis and section 5 summarizes and discusses the results. Section 6 concludes by summarizing the contributions, limitations, and direction for future research.

## 2. THEORETICAL BACKGROUND

### *2.1. Student's Mental Health Illness*

A mental health illness is defined as a health issue that affects the well-being and how a person feels, thinks, behaves, and communicates with others [20,33]. As per American Psychiatric Association, mental health illness is emotional and behavioral or a combination of both types of a health condition that is associated with family, social, or work-related problems [27]. It can be further said that mental health illness is a health issue that affects the emotional and behavioral wellbeing of a person further leading to physiological effects [15]. It has been shown in a study by Kitzrow et al. [14] that mental health illness is further classified as anxiety disorder, depressive disorder, and Schizophrenia disorder. Most of these are exhibited by both adults and teenagers in various age groups [21]. This study focuses mainly on the mental health of college students. In the study by Beiter et al. [5], they showed that college students suffer from anxiety disorder and depressive disorder, and these illnesses are highly related to their social, work, and family issues. In this study, we take into consideration two illnesses that lead to student's mental health problems. Following the research by Hunt and Eisenberg [13], we take into account anxiety disorder and depressive disorder as two types of mental illness among college students. Though Schizophrenia disorders are equally important compared to other three disorders, it is found rarely present among college students [28].

### *2.2. Anxiety Disorder*

As per American Psychiatric Association (APA), anxiety involves the feeling of nervousness, anxiousness, and excessive fear [27]. These feelings are followed by physiological symptoms which persist and appear cyclically over a certain time when anxiety triggers back [8]. Anxiety disorders are mainly classified into three types, Generalized Anxiety Disorder (GAD), panic disorder, and social anxiety disorder. Though APA classification defines few more types, in this study we have confined our research to these three anxiety types. Students with GAD experience severe anxiety or stress about things that are social and part of everyday life like personal safety, jobs, social interactions, and everyday life events [12,34]. Usually, GAD leads them to avoid or seek reassurance about circumstances where the result is unpredictable and be unnecessarily concerned about things that might not give them a favorable outcome. College students are found to exhibit GAD in their daily activities at various events, most GADs are academia related, but some are because of personal and family issues [12]. Panic disorders are those disorders where an event leads to sudden onset of psychological and physiological reactions in people. These reactions may include irregular pulse, sweating, shaking, shortness of breath, etc. College students do exhibit 19ic disorders on many occasions. Many students who exhibit GAD also are seen to show signs of panic disorder.

Social anxiety disorder develops extreme general fear or anxiety in people about other people's reactions to their actions and attitude in a social setting [2]. People who suffer from social anxiety tend to avoid situations where they get attention or exit may be difficult or embarrassing [2]. Students suffer from some sort of anxiety during their college tenure. The anxiety may be related to academics, work, athletics, or social and personal relationships. Students exhibit social anxiety disorder on many occasions and there exists a lot of literature on this topic. Anxiety disorder seems to be a common phenomenon prevailing among students in their daily activities. In this study, we emphasize these three anxiety disorders that affect student's mental health and lead to illness. The datasets have factors in form of features that contribute to these three anxiety disorders.

### *2.3. Depressive Disorder*

Depressive disorder also known as clinical depression is a mood disorder that causes severe symptoms that affect the way a person feels, thinks, and handles daily activities [20]. This disorder is usually characterized by sadness, loss of interest, feelings of guilt or low self-worth, disturbed sleep, low appetite, feelings of tiredness, and poor concentration [20,35]. This study focuses on two types of depressive disorders; persistent depressive disorder and psychotic depression [20]. Persistent depressive disorder, also known as dysthymia is a state where a person's depressive state in form of a depressed mood lasts for at least two years [20]. A person who has been diagnosed with persistent depressive disorder may have major depressive episodes along with periods of less severe symptoms, the cycle of these symptoms lasts for two years, making this disorder more chronic. A large number of students usually suffer from persistent depressive disorder, though there is a likely chance that consistent stress and anxiety leads to this depression making it undiagnosable at an early stage [30]. Psychotic depression differs from persistent depressive disorder, where a person has severe depression in form of psychosis, where symptoms include having disturbing false fixed beliefs and an imaginary sense of hearing or seeing things that others are unable to hear or see. Students do exhibit symptoms of psychotic depression, though such cases are not in large numbers. The severity of psychotic depression makes it more vulnerable to students. In this study, we examine data with factors leading to depressive disorder in general among college students. We try to examine the performance of each ML model in significantly analyzing the data and giving us the best model fit.

### *2.4. Factors of Mental Health Illness*

In this study, we have taken into consideration twelve factors of mental health illness based on previous literature. As shown in figure 1 we have included lack of social support, homesickness, financial problem, gender bias, learning environment, family issues, LGBT, peer relationships, childhood adversities, race, alcohol consumption, and internet addiction as the factors that lead to anxiety and depressive disorder. Generally, the factors responsible for mental health illness are based on the social environment, socioeconomic status, and biological factors [1]. In this study, we have taken into consideration factors related to social environments like social support, learning environment, and peer relationship, and socioeconomic statuses like financial problems as well as factors from demographic characters like race, minority status (LGBT), and Gender. We took into consideration a few independent factors like homesickness, family issues, and childhood adversities which are issues related to the family background of a student. Personal characteristics like internet addiction and alcohol consumption are also considered in this study. In previous literature, it has been shown that social environment, socioeconomic status, demographic characters, family background, and personal characteristics are significantly related to anxiety and depressive disorders among college students [32,4,30,1,18,16]. We try to examine the data related to anxiety and depressive disorder through the factors responsible for the mental health illness. We analyze the data and try to explore the best ML model which can predict student's mental illness outcome. These factors are called "Features" throughout the study after taking into consideration the terminology and taxonomy used in ML model performance evaluation.

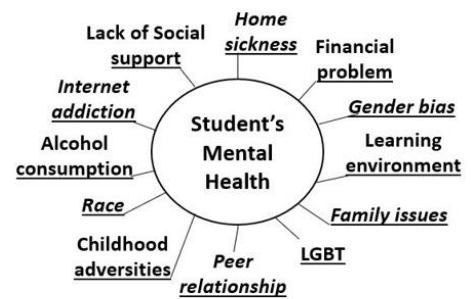

**Figure 1**: Factors of student's mental health illness

### *2.5. Comparative Analysis*

Educational literature is focused on developing educational theories and using empirical analysis in supporting their data [25]. Moreover, few studies have focused on improving the performance of ML models using a single dataset and a specific ML model [23,9]. This has prevented the generalization of the findings. Few studies have considered various ML models, though they have succeeded in their attempt, they have used a single dataset for their purpose [23]. Recent studies have shown that using a single dataset to examine the performance of ML models usually provides limited results in form of either improving feature learning methods or finding an optimal model with the best tradeoff between speed and accuracy [25]. There is a need for a comprehensive study to evaluate and benchmark the performance of various ML models with different educational datasets and map the features of various datasets to appropriate ML models. We aim to extend this research by collectively considering multiple educational datasets, the type of activities being classified, the performance of an expanded portfolio of ML models, and the use of a set of performance metrics to get more insights in understanding the ML models and their relation to educational data in conjunction with the current literature.

## 3. METHODOLOGY

### 3.1. Datasets

We used four educational datasets in a manner that captures the regional diversity of students. We took into consideration four major regions of the USA East Coast, South, West Coast, and Midwest. This is done following the study by Holland et al. [11], where regional students' health policies make each region unique in its perspective. The four datasets are from four public universities located in each region with 20 thousand plus student enrollment. Universities are located in suburban and rural (college town) settings with a metropolitan city within a 20-mile radius. This makes these four datasets unique and appropriate for utilizing them to benchmark various ML models. Table 1 presents the data sets and their features. All the datasets have common features measured on the same scale. The surveys were conducted individually by each institution. In table 1 each feature has items on a 7-point scale. Though the number of items for each feature differs from others, the 7-point scale is commonly maintained for them. Cronbach's alpha is used to measure the internal validity of those items. All features have Cronbach's alpha above 0.8 indicating that the items representing each feature are internally valid. Each dataset had 320 participants, who answered all questions (items) in the survey. A total of 1280 responses were analyzed using ML models.

| No | Feature | No. of Items | Scale | Cronbach's α |
|---|---|---|---|---|
| 1 | Lack of social support (SS) | 3 | 7-point | 0.86 |
| 2 | Home sickness (HS) | 4 | " | 0.88 |
| 3 | Financial problem (FP) | 4 | " | 0.92 |
| 4 | Gender bias (GB) | 4 | " | 0.94 |
| 5 | Learning environment (LE) | 4 | " | 0.87 |
| 6 | Family issues (FI) | 3 | " | 0.94 |
| 7 | LGBT (L) | 3 | " | 0.82 |
| 8 | Peer relationship (PR) | 4 | " | 0.93 |
| 9 | Childhood adversities (CA) | 4 | " | 0.94 |
| 10 | Race (R) | 4 | " | 0.87 |
| 11 | Alcohol consumption (AC) | 3 | " | 0.83 |
| 12 | Internet addiction (IA) | 3 | " | 0.85 |
|  | Outcome variable |  |  |  |
| 1 | Student's mental health illness | 1 | Nominal | - |

**Table 1**: Features obtained from educational datasets

## 4. ANALYSIS

### 4.1. ML Model Evaluation Metrics

We compared different ML models using four educational datasets across various ML performance metrics for evaluating each model. The ML models are Naïve Bayes, Support Vector Machine (SVM) with linear kernel, K-Nearest Neighbor (KNN), Logistic Regression, Stochastic Gradient Descent (SGD), Decision Tree, Decision Tree with entropy, Random Forest, Gradient Boosting Decision Tree (XGBoost), and NGBoost algorithm. We considered deep learning techniques such as neural network-based algorithms, but we concluded that the limited sample size of the datasets would underfit the model and therefore we excluded them from this study [10]. Accuracy is found to be the most popular ML performance metric in most datasets [10], we examined additional metrics such as precision, recall, F1 score, and runtime to facilitate the analysis. Wherein, Recall refers to the rate of correctly classified positives observations among all positives observations, Precision refers to the rate of correctly classified positives among all examples classified as positive, and F1 score represents the harmonic mean of recall and precision [31,32]. Accuracy represents percentage of correct predictions. Table 2 shows the description of ML model metrics employed for evaluating the ML model performance. The four datasets are minimally preprocessed by addressing the missing values and excluding the data (rows of missing values) during the preprocessing stage based on standardizing the format of the data that it would be easier to implement the ML models and interpret the results obtained from the analysis.

| ML Metrics | Definition | Formula |
|---|---|---|
| Accuracy | The ratio of number of correct predictions to the total number of predictions | [TP+TN] / [TP+TN+FP+FN] |
| Precision | The ratio of number of correctly predicted positive values to the total predicted positive values | TP / [TP+FP] |
| Recall | The ratio of correctly predicted positive values to the total number of positive values | TP / [TP+FN] |
| F1 score | Harmonic mean of precision and recall | 2*[Precision*Recall] / [Precision + Recall] |
| Predicting Run Time | Time taken for target classification using test data | |

**Table 2**: ML Model metrics [10,22,28]

The data is standardized in a manner such that each row represents the sample values for each feature (variable) for a specific dataset. In the next step, we split the data into training (80%) and test data (20%) for each dataset after standardizing the format of each dataset with minimal preprocessing. Once the datasets are split into training and test data, we train the various ML models using hyperparameter tuning for each dataset. Each ML model is evaluated with the performance metrics (table 2) on each dataset. Table 2 presents the definitions and formulas for five ML performance metrics. We included accuracy, precision, recall, F1 score, and predicted run time as five metrics for performance evaluation. A potential tradeoff between classification performance and prediction runtime is taken into consideration in this study [10]. We have taken into consideration the concept of Pareto efficiency where no individual criterion can be declared better without a sacrifice in one of the other criteria [10]. This theory is highly applicable in this study as a tradeoff occurs between classification performance accuracy and predicted run time of models.

*4.2. Classification Performance of ML Models*

To examine and evaluate the performance of the various ML models using the various datasets, we considered individual features from the datasets. The outcome variable was represented as a categorical variable where each category represented the status of a student's mental illness. This led to two groups one with mental illness and the other without any. This type of grouping is very common in comparative analysis for understanding each feature and the respective effect of the ML model. Though it allows the researchers to use the number of features to examine the performance of ML models, it is limited in its ability to examine feature importance and feature selection. We continued our analysis with the most significant ML model to identify important features which influence the outcome. We used the Weka software tool to analyze and compare multiple ML models. The Weka software toolkit calculates all the performance metrics and prediction accuracy of models are compared.

*4.3. Feature Importance and Feature Selection*

To take this study further and investigate the influence of features on the outcome variable. We analyzed the datasets using the high-performing ML model to further explore the feature importance in combined datasets. Feature importance is defined as the relevance of features for significantly predicting the outcome variable for better model performance [7]. Relevance of features for better model performance, referred to as feature importance is a product of the field of interpretable machine learning (IML), where important features are derived from the best model fit [22]. Feature importance techniques are of two types permutation feature importance (PFI) and conditional feature importance (CFI) [7]. PFI is based on replacing the feature of interest $X_j$ with a version sampled from the marginal distribution $P(X_j)$ while CFI processes $X_j$ such that the conditional distribution for the set r of remaining features $P(X_j|X_r)$ is preserved [19]. We have used PFI exclusively in this study to examine feature importance. Python Jupyter Notebook was used to analyze the data for feature importance. In this study features are the factors (variables) that are responsible for student's mental health illness as displayed in figure 1.

## 5. RESULTS

*5.1. Descriptive Statistics*

Table 3 shows the means and standard deviations for all the features from the four datasets. Means for all the features are above the scale mid-point of 4. One sample t-tests were conducted to test if one can conclude that scores of the features are above the scale mid-point in the larger population of university students based on the sample means. The results show that all the t-values are statistically significant at the 1 percent level. Thus, we conclude that, in general, the study population finds the sample data to be both valuable and credible. Table 4 (Appendix) is a Pearson's correlation matrix; we can infer that no two features are highly and significantly correlated to impact the results of classification performance of ML models and feature importance.

|  |  | **Descriptive statistics** |  | **One sample t-test** |  |
|---|---|---|---|---|---|
| **No.** | **Feature (N = 1280)** | **Mean** | **SD** | **t** | **p** |
| 1 | Lack of social support (SS) | 4.84 | 1.62 | 18.55 | 0.00 |
| 2 | Home sickness (HS) | 5.28 | 1.44 | 6.96 | 0.00 |
| 3 | Financial problem (FP) | 5.62 | 1.32 | 16.80 | 0.00 |
| 4 | Gender bias (GB) | 6.01 | 1.24 | 29.14 | 0.00 |
| 5 | Learning environment (LE) | 4.88 | 1.68 | 18.74 | 0.00 |
| 6 | Family issues (FI) | 5.34 | 1.12 | 10.86 | 0.00 |
| 7 | LGBT (L) | 4.22 | 1.84 | 4.28 | 0.00 |
| 8 | Peer relationship (PR) | 6.54 | 1.22 | 15.83 | 0.00 |
| 9 | Childhood adversities (CA) | 5.66 | 1.46 | 16.17 | 0.00 |
| 10 | Race (R) | 5.74 | 1.22 | 7.04 | 0.00 |
| 11 | Alcohol consumption (AC) | 4.32 | 1.68 | 6.81 | 0.00 |
| 12 | Internet addiction (IA) | 5.31 | 1.36 | 8.15 | 0.00 |

**Table 3**: Descriptive statistics

## *5.2. Classification Performance of ML Models*

Table 5 depicts the performance of the various ML models on the four datasets. We used five ML metrics to evaluate the classification performance of each model. Accuracy, precision, recall, F1 score, and predicted run time are used as the ML metrics to evaluate model performance. Considering the first metrics, accuracy, we conclude that XGBoost (Extreme gradient boosting) is the best performing model for all the datasets as the value exceeds 0.99 compared to other ML models. The other models are in this order for their performance KNN (k-nearest neighbors), RT (Random Forest), SVM (Support vector machines), DT (Decision tree), Naïve Bayes, SGD (Stochastic gradient descent), Logistic regression, and NGBoost (Natural gradient boosting). Accuracy is considered as the precise indicator of classification performance and from the results (table 4) we infer that XGBoost has the most significant model fit for the four datasets. Considering the three metrics of precision, recall, and F1 score we conclude from the results in table 4 that XGBoost is the best performing model for all the datasets. The other models are in the same order for classification performance considering these three metrics.

The predicted run time metric was the last metric that we considered for classification performance. We took into consideration Pareto efficiency to investigate the tradeoff between classification performance and run time prediction. We conclude that XGBoost is the best performing model for the least predicted run-time. The other models are in this order for their performance SGD, DT, RF, Naïve Bayes, XGBoost, NGBoost, and KNN. Overall, we conclude that XGBoost stands out to be the best performing model on five metrics; accuracy, precision, recall, and F1 score, and best-predicted run-time. Considering Pareto efficiency theory and the tradeoff between classification and run time prediction, we in this study have used XGBoost regression for examining the feature importance from the dataset as we believe that accuracy is a better indicator of classification performance than predicted run time.

|  |  | **Naïve Bayes** | **SVM** | **KNN** | **SGD** | **Logistic reg** | **DT** | **RF** | **XGBoost** | **NGBoost** |
|---|---|---|---|---|---|---|---|---|---|---|
| **Accuracy** | D1 | 0.801 | 0.774 | 0.783 | 0.700 | 0.720 | 0.766 | 0.780 | **0.999** | 0.736 |
|  | D2 | 0.779 | 0.776 | 0.780 | 0.730 | 0.740 | 0.775 | 0.795 | **0.999** | 0.781 |
|  | D3 | 0.721 | 0.756 | 0.778 | 0.729 | 0.738 | 0.711 | 0.739 | **0.994** | 0.775 |
|  | D4 | 0.768 | 0.772 | 0.763 | 0.766 | 0.768 | 0.736 | 0.748 | **0.993** | 0.775 |
| **Precision** | D1 | 0.710 | 1 | 0.795 | 0.723 | 0.727 | 0.784 | 0.782 | **1** | 0.731 |
|  | D2 | 0.780 | 1 | 1 | 0.720 | 0.704 | 0.780 | 1 | **1** | 0.714 |
|  | D3 | 0.704 | 0.790 | 0.783 | 0.703 | 0.712 | 0.763 | 0.772 | **0.993** | 0.768 |
|  | D4 | 0.712 | 0.794 | 0.778 | 0.723 | 0.724 | 0.787 | 0.786 | **1** | 0.746 |
| **Recall** | D1 | 0.714 | 1 | 1 | 0.703 | 0.727 | 0.773 | 1 | **1** | 0.738 |
|  | D2 | 0.720 | 0.770 | 0.79 | 0.730 | 0.743 | 0.778 | 0.740 | **1** | 0.746 |
|  | D3 | 0.716 | 0.763 | 0.79 | 0.710 | 0.721 | 0.743 | 0.736 | **0.992** | 0.748 |
|  | D4 | 0.726 | 0.776 | 1 | 0.728 | 0.741 | 0.778 | 0.757 | **1** | 0.767 |
| **F1 score** | D1 | 0.732 | 0.786 | 1 | 0.737 | 0.748 | 0.763 | 0.748 | **0.998** | 0.748 |
|  | D2 | 0.718 | 0.748 | 0.786 | 0.724 | 0.736 | 0.738 | 0.718 | **0.983** | 0.717 |
|  | D3 | 0.722 | 0.772 | 0.792 | 0.748 | 0.744 | 0.743 | 0.743 | **1** | 0.763 |
|  | D4 | 0.768 | 0.774 | 0.797 | 0.762 | 0.761 | 0.758 | 0.753 | **1** | 0.768 |
| **Predicting Run Time [s]** | D1 | 6.381 | 652.3 | 14,863 | 0.168 | 8.824 | 0.186 | 1.653 | **0.140** | 404.768 |
|  | D2 | 4.366 | 332.5 | 7852 | 0.328 | 10.624 | 0.283 | 3.614 | **0.316** | 854.634 |
|  | D3 | 1.488 | 142.5 | 4885 | 0.234 | 6.625 | 0.268 | 1.262 | **0.230** | 122.888 |
|  | D4 | 1.383 | 36.6 | 3863 | 0.198 | 6.183 | 0.166 | 1.183 | **0.126** | 116.826 |

**Table 5**: Performance matrix for ML models

*5.3. Feature Importance of Dataset Features*

As the next step in our research, we have examined the individual effects of features from the dataset. From our model comparison stage, we concluded that XGBoost is the best classification performance model when it comes to accuracy, precision, recall, and F1 score and the least predicted run-time. To examine the feature importance, we combined and analyzed the four datasets using XGBoost regression with the scikit-learn package. The feature importance was computed using three methods; permutation method in XGBoost regressor, built-in feature importance in XGBoost algorithm, and SHAP value computation. We used permutation_importance from scikit-learn on XGBoost to run the regression because XGBoost implements the scikit-learn interface API. The permutation method randomly shuffles each feature and computes the change in the model's performance for the XGBoost model. The features that impact the performance the most are the most important ones to influence the model. Figure 2 displays feature importance through the permutation method applied to the XGBoost regressor. The top four features include lack of social support (SS), learning environment (LE), childhood adversities (CA), and alcohol consumption (AC). We use concurrent validity as a method to verify these four features by comparing them to the results from the other two methods SHAP value computation and built-in feature importance in the XGBoost algorithm.

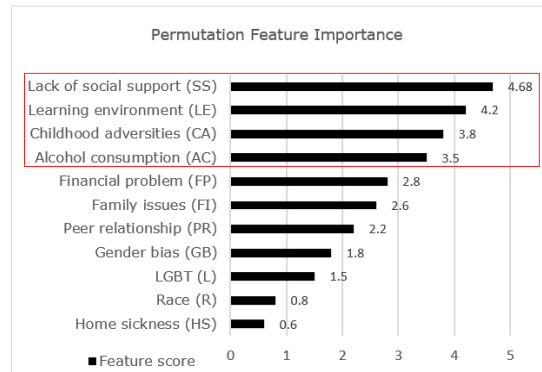

Figure 2: Permutation feature importance in XGBoost regressor

In the second method, we compute the feature importance using built-in feature importance in the XGBoost algorithm. We then try to examine if the features are ranked in a similar hierarchy compared to that of the permutation method. Figure 3 displays feature importance through built-in feature importance in the XGBoost algorithm. The top four features include lack of social support (SS), learning environment (LE), alcohol consumption (AC), and childhood adversities (CA). It noticeable that the third and the fourth position have interchanged.

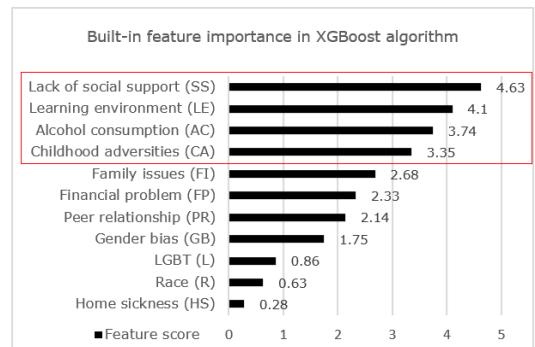

Figure 3: Built-in feature importance in XGBoost algorithm

In the third method we used SHAP package to compute feature importance in XGBoost. This method uses the Shapley values from game theory to estimate the feature contribution in outcome prediction. Figure 4 displays feature importance through SHAP values applied to the XGBoost regressor. The top four features include lack of social support (SS), learning environment (LE), childhood adversities (CA), and alcohol consumption (AC). It is clear from this analysis that four features are consistently at the top with only one variation in the built-in feature importance method. We consider the feature ranking variation as negligent and non-effective to We conclude that these four features; lack of social support (SS), learning environment (LE), childhood adversities (CA), and alcohol consumption (AC) are significantly responsible for causing mental health illness among college students taking into account the three methods of feature importance.

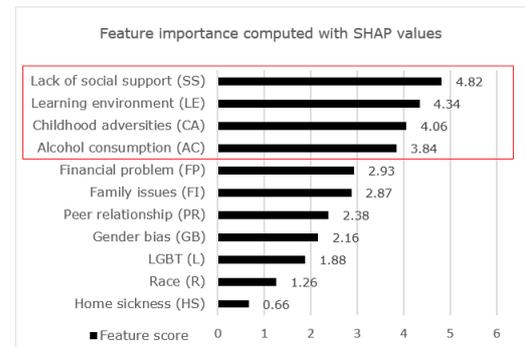

Figure 4: Feature importance computed using SHAP values

## 6. DISCUSSION AND IMPLICATION

This study examined two goals, first, it explored the best performing ML model for classifying educational data and second it analyzed the importance of each feature from the datasets. Results show that XGBoost is the best performing classification model suitable for an educational dataset to examine the outcome variable of students' mental illness. In previous literature, tree-based algorithms like XGBoost have been found very effective in classifying sensor data in the field of human activity recognition (HAR) data [17]. This study attests to the claims of previous studies by exploring the application of the tree-based XGBoost algorithm in the educational dataset. Moreover, it takes into consideration the relative importance of features and their role in the ML model's performance. After investigating the dataset using XGBoost regressor with the help of three methods we conclude that four features; lack of social support (SS), learning environment (LE), childhood adversities (CA), and alcohol consumption (AC) are significantly responsible for affecting a student's mental health, further resulting in continued illness.

### *6.1. Theoretical Implications*

From a theoretical point of view, this research supports previous literature on the tree-based algorithm by making two contributions. First, it introduces the application of the XGBoost model to an educational dataset primarily composed of survey-based data. Second, it examines the feature importance using three methods applied to the XGBoost regressor. This study took into account multiple ML metrics going beyond few traditional metrics like accuracy and F1 score [7] and found XGBoost to the best classification performing model. All ML models performed well on four metrics; accuracy, precision, recall, and F1 score, while predicted run-time was used as an additional metrics to measure the run-time speed of models where only one model; XGBoost was found to be the best performing model. Although, as per previous literature these observations were expected for ML model performance on traditional metrics (Accuracy & F1 score), we theoretically contributed by adding three more metrics like precision, recall, and predicted run-time to the comparative analysis.

Only few studies in education literature have studied the comparison of ML models and have found model accuracies of 0.95 and F1 score of 0.86 with a single dataset [7], we in this study have obtained much higher accuracies and F1 scores with a large range of ML models. Feature importance has been a unique exploration completed in this study for an educational dataset. This was done to better understand which features (factors/variables) significantly contribute to the outcome variable. Three methods were used as an extension to the best performing XGBoost to explore feature importance taking into consideration the importance of concurrent validity. We theoretically contribute to the science of mental health problems and clinical psychology by exploring four significant features contributing to student's health illnesses. This is a unique undertaking done in this study which gives it an edge as it contributes methodologically (ML models) as well as theoretically to the science of educational studies.

### *6.2. Industrial Implications*

From an industrial point of view, this study's findings are helpful to ML AI (machine-learning artificial intelligence) application developers. We ran predicted run time as an ML metric to measure and compare the classification performance of models. After accuracy, predicted run time is an important indicator of model performance [7], especially if the model is supposed to analyze a large set of real-time data in the dynamic software application. We found XGBoost to be the better performing model when it came to predicting run time. For a real-time application that requires analyzing educational data, the main concern for designing the application would be minimizing prediction run time while maintaining acceptable classification performance. In this case, XGBoost would be a better model than other models.

## 7. LIMITATIONS AND FUTURE WORK

This study has limitations that future studies could address. First, the sample size that we acquired is not sufficient to make a generalized conclusion. However, due to the widely held notion of efficient and complete sampling of survey-based behavioral data, we can support a conclusive argument that XGBoost is a better performing classification model for these datasets. Researchers should explore larger datasets with diverse features to generalize their findings. Second, we have used Pareto efficient to investigate the tradeoff between accuracy and predicted run-time, but have based on conclusion on accuracy to determine the best performing model. Researchers should examine diverse datasets to check if more metrics can influence any tradeoff and should support their findings theoretically and empirically.

**APPENDIX**

| Feature | SS | HS | FP | GB | LE | FI | L | PR | CA | R | AC | IA |
|---|---|---|---|---|---|---|---|---|---|---|---|---|
| **SS** | 1 | | | | | | | | | | | |
| **HS** | 0.21 | 1 | | | | | | | | | | |
| **FP** | 0.28 | 0.18 | 1 | | | | | | | | | |
| **GB** | 0.51 | 0.23 | 0.52 | 1 | | | | | | | | |
| **LE** | 0.68 | 0.42 | 0.27** | 0.46 | 1 | | | | | | | |
| **FI** | 0.33 | 0.56 | 0.16 | 0.18 | 0.51 | 1 | | | | | | |
| **L** | 0.46* | 0.27 | 0.17 | 0.42 | 0.14 | 0.38 | 1 | | | | | |
| **PR** | 0.32 | 0.16 | 0.33 | 0.18 | 0.42 | 0.46 | 0.27 | 1 | | | | |
| **CA** | 0.42* | 0.17 | 0.18 | 0.23* | 0.21* | 0.33 | 0.42 | 0.38 | 1 | | | |
| **R** | 0.38 | 0.52 | 0.16** | 0.27 | 0.46 | 0.21 | 0.18 | 0.46 | 0.14* | 1 | | |
| **AC** | 0.63 | 0.21 | 0.23 | 0.42 | 0.21 | 0.16 | 0.46 | 0.16 | 0.28 | 0.42 | 1 | |
| **IA** | 0.58 | 0.46 | 0.21 | 0.14 | 0.18* | 0.51 | 0.16 | 0.28 | 0.33 | 0.23 | 0.38 | 1 |

**p<0.01, *p<0.05

**Table 4**: Correlation matrix